**Stability evaluation of the Russian sociologists' online community: 2011-2018 years[1]**

*Aryuna Kim[a] and Daria Maltseva[b]*

[a] Research Assistant at the International laboratory for Applied Network Research, HSE University, Moscow, Russia (avkim@hse.ru), ORCID: 0000-0002-3119-1087, https://www.facebook.com/aryungutang

[b] Deputy Head at the International laboratory for Applied Network Research, HSE University, Moscow, Russia (dmalceva@hse.ru), ORCID: 0000-0003-1789-1711, https://www.facebook.com/darya.maltseva

## Abstract

This study deals with the stability evaluation of the online community of Russian sociologists. Based on the data from the Facebook group, which consists of 7 years of communication from 2011 till 2018, we constructed the networks based on commenting and reacting. The participants' activity includes four main periods for stability evaluation of the community. The blockmodeling discloses the structural patterns of interactions in the community. The results show the "core-periphery" type of the global structure. The core and periphery are structured differently in the networks of comments and reactions. The stability between the positions in the global structure is high and while the structure may vary in some periods, the size of the core and periphery fluctuates. However, the stability within the positions of the global structure is low, according to the modified Rand index (Cugmas, Ferligoj, 2018).

---

**Introduction**

Along with the development of society in the 20th century, the concept of community, formulated in classical sociology, began to denote new forms of interaction between individuals, such as those appearing in online communities (Rheingold 1993, 2000). The concept of communities of practice (Lave and Wenger, 1991) was proposed to denote professional online communities, where the same practice, knowledge and identity are shared among professional groups on the Internet. Studies have shown that the motivation for participation in such communities may vary (Hew, Hara, 2007; Hur, Brush, 2009), which leads to different structural forms of interaction within the community (Faust, Wasserman, 1992). The research (Kronegger et al., 2011; Rykov, 2016; Schenkel et al., 2001) has shown that the structure in such communities has the form of "core-periphery", where the core consists of members who are highly connected with each other, and the periphery is linked only with the core and not between themselves. However, the question remains how stable these global structures are, do they change in time, and how the individual members move between positions in these structures.

In our study, we consider the structure of communication based on the largest Facebook group of Russian sociologists. We believe this group can be seen as an example of a community of practice, as it brings together people with the same professional interests in sociological research and are actively involved in professional discussions. From the epistemological perspective, the case of Russian sociologists is important to study because the sociological discipline in the USSR and modern Russia has a unique and difficult history of formation. The description of the history of Russian sociology is problematic and "defies rational description at all" because of too many different facts, turning any "beautiful and exhaustive historiographical scheme into an arbitrary construction" (Batigin, Deviatko, 1994). The history of this discipline influenced the formation of the corresponding research community. The empirical studies, including those using a structural perspective, have shown that there are different groups of researchers that can be found within the "offline" sociologists' community (Sokolov et al., 2010; Batigin, Gradoselskaya, 2001). However, an analysis of communities in an online format, by its nature suggesting more horizontal relationships, can bring different results. While there are some studies of online communities of sociologists (Maltseva, 2016; Barkhatova, 2020), to the best of our knowledge, this study is the first attempt to make a comprehensive overview not only of the community's global structure, but also its stability between and within its subgroups over a long-term period. Based on the information on posting and commenting, we observe the structural characteristics of a large professional Facebook group over 7 years (2011 -- 2018) using the structural perspective and methods of social network analysis (SNA).

The structure of the article is as follows. The Literature review describes the theoretical background of studying communities in sociological research and identifies some characteristics of the online community of practice. It also presents previous studies on the Russian sociological community, giving special emphasis to the research done through the structural perspective and in online communication. The Data and methodology section characterizes our case study and describes the selected online professional community, presents data collection and network construction processes, and describes the methodology used for the analysis. The Results section provides the main findings: the global structure of the community, and its stability between and within the obtained groups. The article finishes with Discussion and Conclusion.

## Literature review

### Communities: theoretical background

The concept of community has played an important role in the theory construction in social sciences. In the 19[th] century, it was defined with clear ideological and political consequences (in works of Gobbs, Spinoza, Lock, Kant, Fichte, and Tonnies). Its further development relates to Marx, Engels, Durkheim, Weber, and Simmel, who studied the economic, social and cultural processes of urbanization and their impact on the social alienation, class formation, and the creation and destruction of individual identities.

Study of the industrialization impact on the preservation of urban communities continued by the Chicago school of sociology within the areas of urban sociology and the sociology of communities. As part of the field research in the 1920-1930s, the Chicago school confirmed earlier development of the community studies. The data showed that instead of being included in a separate cohesive community, urban residents are limited members of various, loosely connected and limited social networks. Such weak and disorganized relationships cannot provide social support to their members, making individuals more dependent on formal organizations, such as employment agencies. Consequently, indirect secondary relations tied urban residents to the city. Evidence of the loss of solidarity and the effects of disorganization was found in areas as diverse as collective action, crime and migration (Wirth, 1938).

Canadian sociologist Wellman suggested considering the Community Question from the structural position of SNA. He raised the question of how large-scale social system divisions of labor associated with urbanization and industrialization affect the organization and content of the main primary relations (Wellman, 1979). As the division of labor in industrial bureaucratic societies has weakened solidarity in communities, the findings of Chicago school were labeled by the term "Lost community". However, the mistake of the representatives of the Lost community approach was that "because of its assumption that strong primary ties naturally occur only in densely knit, self-contained solidarities, the argument has unduly neglected the question of whether primary ties have been structurally transformed, rather than attenuated, in industrial bureaucratic social systems" (Wellman, 1979: 1205).

A reaction of many urban sociologists to the evidence of the "Lost community" was the development of the opposite approach, which claimed that neighboring and related solidarity groups continue to exist successfully in industrial bureaucratic social systems (Wellman, 1979: 1205). Field research, conducted in 1940-60s, showed that citizens continue to organize personal communities in homogeneous living and working spaces (on the scale of the neighborhood, their friends, and work). The approach of the "Saved community" looks more positively at people's ability to adapt to complex social conditions; even in complex social and economic environments, people seek to organize social structures of mutual support. Among the examples supporting this point of view are the social movements of citizens protesting the reconstruction of urban neighborhoods (to preserve the existing community).

According to Wellman, a common problem for the two approaches is that in many studies, the Community Question included two components 1) the submission of a normative nature on the solidarity of sentiment in a community, and 2) an awareness of the specific spatial distribution of major linkages in local areas. "As a result of this confounding, the fundamentally structural Community Question has often been transmuted into a search for local solidarity, rather than a search for functioning primary ties, wherever located and however solidary" (Wellman, 1979: 1202). As such, the locality can no longer be considered as one of its main constitutive characteristics of the communities.

Further development of technologies contributed to the confirmation of this idea, but at the same time raised new Community Questions. Whereas the main volume of sociological community studies in the 20th century sought the answers to the questions posed by scientists of the 19th century, drastic revolutionary changes in technologies of the 1990-2000s meant new challenges to communities in their traditional forms. For the first time the concept of online, or virtual, community was presented by Rheingold (1993, 2000), who described one of the first communities existing in network form – The WELL (Whole Earth Electronic Link). This study was one of the first talking about the existence of communities in the virtual environments. It has shown that members of online communities, combined with each other's interests, work or training, reflect the same kind of characteristics that can be found in offline communities, such as, for example, the formation of a common language, rules of conduct and compliance, social support, and the creation of a common shared history. Virtual communities are "social associations that arise from the network, when a sufficient number of people lead public discussions long enough, with enough human feeling, to form networks of personal relationships in cyberspace" (Rheingold, 1993). By the mid-1990s, due to the spread of personal computers and the expansion of the Internet, real and virtual life began to converge. Cyberspace with its virtual communities, formerly understood as separate spheres of life, gradually began to enter into people's daily practices. Studies of the intersection and complementarity of online practices with everyday practices have concluded that Internet-mediated communication has become another tool in the overall system of communication.

Conceptualizing the community as a social network, Wellman et al. (2002: 153) defined the "community before the Internet" as a homogeneous group with neighborhood interaction; as "networks of interpersonal ties that provide sociability, support, information, a sense of belonging, and social identity". Comparing the online community with a neighborhood, having some local geographical patterns, the authors denoted the new feature of online communication as "network individualism": "In networked societies, boundaries are more permeable, interactions are with diverse others, linkages switch between multiple networks, and hierarchies are flatter and more recursive" (Wellman et al. 2002: 160).

Currently, online form of communication is typical for various types of communities, including professional communities. Traditionally, these communities were studied in sociology of profession and professional groups, based on the division of labor presented in the classical works of Spencer, Marks, Durkheim, and Weber. The technological progress influenced the appearance of new forms of studying professions. To study professional communities in an online environment, the concept of community of practice (CoP), proposed in 1991 by Lave and Wenger (1991), may be relevant. CoP is defined as "groups of people who share a concern, a set of problems, or a passion about a topic, and who deepen their knowledge and expertise in this area by interacting on an ongoing basis" (Wenger et al., 2002: 4). This term was developed in the context of the study of traditional apprenticeships. Describing the history of professional groups, Durkheim argued that the professional community could provide social connections that are important for strengthening the matter of social trust and mutual commitment, even when the forces of industrialization and accompanying social disruptions are trying to break the historical ties that unite people in the villages (Wenger et al., 2002). The structural model of the CoP presupposes the existence of three main elements – the community, the practice, and the sphere of interests. To these three characteristics, the technology can be added, as the use of such means of communication became part of the CoP in online platforms. Hence, special digital habitat of CoP exists -- a virtual settlement (Wenger et al., 2002).

Various studies have shown the positive effect of Internet use in the communication and participation in the offline professional communities (Renninger, Shumar 2002; Schlager at al., 2002). Researchers have distinguished different motivations for participating in communities. Hew and Hara (2007) list four main reasons to share knowledge in the teachers community: (a) collectivism: teachers share knowledge to improve the welfare of community members, (b) reciprocity: teachers have received help from others and want to give it back,

(c) personal gain: sharing knowledge, teachers gain new knowledge, and (d) altruism: teachers feel empathy for other teachers' struggles and would like to support them. Hur and Brush (2009) found the following reasons to participate in teacher's online communities: (a) sharing emotions, (b) utilizing the advantages of online environments, (c) combating teacher and isolation, (d) exploring ideas, and (e) experiencing a sense of camaraderie. Some types of these reasons could be found in other types of professional communities.

The a-symmetric activity of participants in online professional communities is another interest of the researchers working in the community studies (Nonnecke, Preece 2003; Rafaeli et al., 2004). In CoP, different roles are assigned to their members according to their participation in the community, such as newcomers, insiders, or outsiders. The model of participants' entry into communication in the online community is based on the principle of acceptable peripheral participation (Lave and Wenger, 1991). The model includes five trajectories of participants' communication in the community:

- peripheral trajectory - for observers (Peripheral-Lurker) who do not contribute to the community resources;
- entries trajectory - for newcomers (Inbound-Novice) who are trying to make some contribution and to community activities;
- internal trajectory - for regular members (Insider-Regular) who are actively involved in the community's activities;
- borderline trajectory - for leaders who interact with participants, express themselves in conceptual ideas, and correct problems of interaction in the community;
- alienations trajectory - for participants leaving the community (Outbound-Elder), no matter for what reason, but switched to another activity or left for another community.

According to the ideas of Lave and Wenger, the participants could move from legitimate peripheral participation to full participation in the community.

While communication in online professional communities is crucial for a participant's career improvement and overall community's development, many people prefer lurking. It was shown that there is "passive attention over active participation" (Rafaeli et al., 2004: 1). The practice of lurking is normal role for overall community's communication and should not be considered as dysfunction (Nonnecke, Preece 2003). Authors described several reasons of lurking such as to ensure privacy, or being shy about posting, or leaving the group. Based on the study of Rafaeli et al. (2004), the reported proportion of lurkers varies from 90% to 50% of the whole professional online community. To measure the relation between lurking, de-lurking, social capital, and cultural capital in online communities, researchers analyzed 82 most active online forums from online courses. The results showed that there is a correlation between active participation and virtual social capital, while activity levels no longer correlate with virtual social capital as the number of posting is increasing, and virtual social capital positively correlates with the levels of de-lurking in the community.

Therefore, differences in the motivation and activity of community members can lead to different structural characteristics of communities. Structural analysis shows the presence of various types of structures in communities. Faust and Wasserman (1992) describe five types of structures that display certain properties of communities: cohesive subgroups, core - periphery, centralized, hierarchical, and transitive structures. Cohesive subgroups are not connected with each other. In core-periphery or center-periphery structures, one group is defined as a "core group", which members are highly linked with each other, and second group defined as a "peripheral group", where its members are linked with the members of core group, but not with each other. In a centralized structure, all relationships are going from one group member. In a hierarchy, the relational ties are directed from each member "below" to another one immediately "above" it. Transitive structure is characterized by the principle: if A is connected to B, and B is connected to C, then A is also connected to C. This structure is "similar to a hierarchy, but all interposition ties that are implied by the property of transitivity are also present" (Faust, Wasserman, 1992: 41). It is important to note that structure is a theoretical construct, since real empirical network data can consist of variations of data from different structural patterns. Researchers found that professional communities and CoP can be characterized by the core-periphery structural type (Kronegger et al., 2011; Rykov, 2016; Schenkel et al., 2001), consisting of two main groups: participants who communicate highly among each other and those who interact only with the first group, but not among each other.

**Russian sociological community: development and previous studies**

The history of sociology in the USSR and modern Russia, and the corresponding community, can be described as nonlinear and dramatic (Batigin Deviatko, 1994). It is usually divided into two unrelated stages: the pre-revolutionary sociology developing before the October Revolution in 1917 and the sociology developing in the Soviet era. After the October revolution, Marxism-Leninism became a state science, and the tasks of sociology were confined to ideological control. It forced some scholars to immigrate: one of the prime examples is Pitirim Sorokin, who left the USSR in the 1920s and became a world renowned sociologist and founder of the first sociological faculty in Harvard University (Firsov, 2012). In the mid-1970s, an undergraduate specialization in applied sociology was set up at the Ural, Minsk, and Leningrad universities. In 1984, the first sociological departments were established in the Moscow and Leningrad State Universities (Titarenko, Zdravomyslova, 2017). Modern Russian sociology was born after the economic reforms of 1990's, followed by the creation of capitalistic relations, which made Russian intellectual elites revise the problems of Russian society and try to find the answers from the Western sociology. In 1988, the first All-Russian Public Opinion Research Center (ARPORC, also VCIOM) was launched (Titarenko, Zdravomyslova, 2017).

Certain aspects of the development of sociology in the USSR and Russia, and the formation of the corresponding academic community, were studied within the framework of historiographical research. In recent decades, many sources appeared that allow researchers to dive into the historical context: documentary evidence about the history of sociology in the USSR and Russia, personal stories based on memoirs, biographies, biographical and thematic interviews of famous sociologists. The history of sociology in the USSR and Russia was analyzed in detail in the works of Firsov and Moskvichev, Osipov (2012; 2008). Kozlova described the history of creation of the first academic Institute of Sociology (2018). Gorshkov (2017) described the overall history of Russian sociology (see all observed studies in the Table 1 in the Appendix). These studies are based on the analysis of documents using the method of historiographical description. However, official documents and protocols often create only an "external, institutional chronology of sociological science" (Batigin, 1999: 5), which is why the combination of different data sources is important.

One of the features of modern Russian sociology is the existence of two groups within the community -- academics, representing more theoretically driven academic research, as well as teaching, and pollsters or practitioners working in the applied commercial sphere. Though these groups partially overlap, most often, the studies referred to either one or the other group of researchers. In these studies, some aspects of communication and collaboration were considered, including those carried out via online platforms.

Based on the in-depth biographical interviews with more than 200 scholars mostly from the academy, Doktorov (2016) described the individual trajectories of academic careers among sociologists. Using the transformed biographical information from the interviews, the collaboration networks of sociologists via network analysis were studied (Batigin,

Gradoselskaya, 2001; Korpachev, 2006; Mazina, 2013; Maltseva, et al., 2017; Maltseva, Moiseev, 2018). The analysis of egocentric networks showed the career paths of sociologists and the development of the whole sociological community (Batigin, Gradoselskaya, 2001). An analysis of affiliation networks of researchers showed organizations, research groups, research centers, and cities, which influenced the development of the sociological community in the 1960s-90s (Korpachev, 2006); different generation groups of sociologists were considered (Mazina, 2013). Based on the same data, the approach for studying the meaning of relations underlying the formation of a professional community of sociologists was proposed (Maltseva, et al., 2017; Maltseva, Moiseev, 2018). A study of the local academic community of sociologists (in St. Petersburg), based on citation and survey analysis (Sokolov et al., 2010), identified three main segments of sociologists: 1) those oriented towards the West, publishing in international academic journals and communicating with colleagues outside of Russia; 2) those publishing only in Russian journals and focused on communication on the national level; 3) "zone in transition", those who do not have a unidirectional strategy of development in Russia. The study of information culture and professional communication in the community of "applied" sociologists, via a survey among research agencies (Zadorin, Maltseva, 2013), showed that Internet communication and social media are important sources of communication for practitioners. An analysis of discussion in the professional online community of sociologists, which includes sociologists from academia and research agencies on a socially significant topic (Maltseva, 2016), identified several main leaders attracting other participants. A recent study (Barkhatova, 2020) illustrated the structural characteristics of communication in the online community of Russian sociologists, where the community consists of a small core group and a huge peripheral group.

Based on this review, we can assume that currently, the community of Russian sociologists can be seen as varying for various reasons, both between groups (academicians and practitioners) and within them (e.g., schools of thought, generations, orientations towards international or national levels). These divisions may increase due to a remarkably high degree of centralization in Russia, formed around Moscow and St. Petersburg, as well as a lack of platforms for the direct communication of sociologists from different groups (own events, journals more oriented toward one or another group, etc.). In this sense, the online professional community and CoP can be the platform for bringing different people together and forming a joint community.

## Data and methodology

**Research questions**

In our research, we analyze the structural characteristics of a particular group of Russian sociologists on Facebook. We formulate the following research questions:

1. Which type of the **global structure** of the studied online community can be identified, according to Faust and Wasserman (1992)? Does this structure fit the **core-periphery model**, as was found in previous studies (Kronegger et al., 2011; Rykov, 2016; Schenkel et al., 2001)?

2. How stable are the patterns of interactions **between the positions** of the global structure, and how do they change in time?

3. How stable are the trajectories of individual membership **within the positions** of the global structure? Can we observe the **migration from one trajectory to another** (e.g., from the periphery to the core, or vice versa), as is proposed in the CoP literature (Lave and Wenger, 1991)?

**The context of the study**

In this study, the data are from the Facebook group "Manufactura Socpokh", whose full name[2] could be translated to English as "Manufactory of sociological adventures". The online group was launched on September 30, 2011, by a sociologist from both professional groups -- academics and practitioners. In this Facebook group, members are defined as a sociologist by their own denomination, and they do not necessarily have a specialized education (which is in line with the peculiarities of the formation of sociology as a science in Russia). In July 2021, it included 3,363 members; it is the largest group representing sociologists on the Russian social media platforms. This is a closed group, where one must request to join the group and be approved by the moderator before being able to see the group's content and take part in the discussions; however, in a certain period (including the period of data collection), the group privacy settings were switched to an open group. Discussions in the group are devoted to empirical and theoretical issues of sociology but do not exclude news feed and thematic disputes.

As in any other Facebook group, there are several possible types of activity between the group members: writing a post, commenting to a post or other comment (the option appeared in 2015), and giving an emotional feedback to a post or comment (the most popular and used is "like", but in 2016 six other options appeared – "love", "wow", "haha", "sad", "angry", "thankful"). We can use the frame of CoP to study this community as it has all the necessary elements: 1) the community itself – group of professional sociological researchers, including academicians and practitioners (pollsters), 2) the joint sphere of interest – the same professional research expertise, field of activity, 3) the practice – research and applied activities of the community members, 4) shared virtual habitat - Facebook group. This group is an interesting case of self-organization of the sociological community representatives, and its originality and uniqueness are due to the following characteristics:

- Long period of existence – since 2011 and up to 2021 (the data available for analysis is up to 2018).

- Diversity of participants representing two main segments of Russian sociology (academics and practitioners), differing by institutions and organizations, age, gender, region of residence, providing opportunities for horizontal communication between the community members.

- Active discussions, attracting community members with divergent viewpoints, compliance with the rules of academic and professional freedom, without any censorship and banning.

---

[2] Мануфактура "СОЦПОХ", or «Социологические похождения» (n.d.). Retrieved from https://www.facebook.com/groups/socpokh/

Even though the structures observed in offline and online worlds are not the same, some similarities between them were claimed to exist (Reich et al. 2012, Subrahmanyam et al. 2008). We fully understand that this group does not represent the community of sociologists in Russia; however, it can be a nice representation of its most active part, present on Facebook.

**Data collection**

The data were collected in January 2018, using Facebook`s official API. The database created from the collected data consists of more than 34,000 posts and comments written from October 2011 up to January 2018 by 818 group members. The collected dataset consisted two parts. In the first, the information on the date, type of publication (post, comment to post, comment to comment), post text, author, achieved reactions, and number of comments (8 types). In the second part, the information on the relationships between any type of publication and author, types of publication including -- post and comment, comment and comment, also reaction and author, types of publication and reaction (4 relations) was stored.

The data was stored in a table in .csv format. Such organization of the database was crucial for the creation of networks.

## Methodology

In accordance with the research goal and main questions, this study uses social network analysis (SNA) as a general methodological approach for revealing structural characteristics of the observed community. SNA analyzes social structures, which determine roles in the communities, and can be widely applied in community research, for instance, to study nonprofit organizational research capacity or community collaborations among professional groups (Johnson, et al., 2010; Korazim-Kőrösy et al., 2014). It is supplemented by some quantitative and qualitative types of analyses, which are widely used for studying communities and their different types, including CoP (MacNair, 2010). Based on the data, we can construct two types of networks, where the group members represent vertices (or actors), and the relations between them are based on commenting to each other (communication) and giving reactions to each other (reaction).

To reveal the global structure of the observed community, we use the blockmodeling approach. "Blockmodeling seeks to cluster units which have substantially similar patterns of relationships with others and interpret the pattern of relationships among the clusters" (Batagelj, et al. 2004: 455). In our case, it allows us to cluster group members according to their similar structural characteristics (interactions with others), to distinguish the social positions of the group members, to identify the relationships among the positions that define roles of each position, and to identify the fundamental network structure, assigning it to one of the types observed by Faust and Wasserman (1992). Blockmodeling is based on the principle of equivalence, which can be structural or regular. Units are structurally equivalent "if they are connected to the rest of the network in identical ways" (Batagelj, et al., 2004: 457), and regularly equivalent "if they are equally connected to equivalent others" (Batagelj, et al. 2004: 457). In our study, we construct blockmodels based on the structural type of equivalence, as we search for the patterns of interactions. We apply an indirect approach to blockmodeling (Batagelj, et al., 2004: 457), as it works better with rather large (several hundreds of nodes) networks.

We study the stability of the obtained structures, as we observe the fundamental network structures in different time periods of group activity. We construct a temporal network, splitting the data into four time periods based on group activity in posting, commenting, and giving reactions, and use the blockmodeling approach to obtain the global structures in each period, and study the stability of patterns of interactions between the positions of the structures. "Clusters of equivalent or similar members in the community are called positions, and the role structure is obtained by links between these positions" (Faust, Wasserman, 1992). To study the stability of the trajectories of individual membership within the positions of the global structure and their change through time, we visualized the trajectories of the community members between the clusters found in the global structure. To evaluate the stability of the group members' trajectories, we use a modified Rand index (Cugmas, Ferlogoj, 2018) – a special index which compares two or more network partitions of non-equal sets of units (as there are new members coming to the community and members leaving the community) and indicates the level of networks' stability. This index depends on the stability of clusters, where the splitting and merging clusters have the same impact on the value of the index. The higher the value of the modified Rand index is, the more similar or stable the partitions are. Low values of the modified Rand index means random and independent partitions.

The analyses described are provided for both types of networks – based on comments and reactions. For the computations, we use the program Pajek[3] (Batagelj, et al. 2004), specially developed for the analysis and visualization of large-scale networks.

**Network construction**

We built networks based on two types of interaction in the online community – commenting posts and comments, and reactions. In the latter network, all types of reactions were merged to one type ("reaction"), however, it should be noted that the "like" reaction is the dominant one. To produce these networks, a program Text2Pajek[4] was used, which transforms the column formed .txt and .csv files to networks[5] to the .net format, which can be directly used for the analysis in Pajek. The formula for the network construction is provided in the Appendix.

Both obtained networks are directed – as one group member is commenting to another one or giving a reaction to their posts and comments. Two networks are also weighted, which means that the strength of ties is based on the interaction frequency among participants (number of comments to one's post or comments, number of giving reactions to a group member).[6] The comments network **CN** consists of 818 vertices, or group members, connected to each other through the relations of commenting on each other's posts or comments. The reactions network **RN** consists of 1,539 vertices, or group members, connected to each other through the relation of giving reactions to each other's posts or comments.

Temporal networks for the four periods (T1, T2, T3, and T4) were constructed manually. The data were split into four parts according to group activity, and then the networks of commenting and reaction for each period (**CN1, CN2, CN3, and CN4; RN1, RN2, RN3, and RN4**) were constructed.

---

[3] Retrieved from http://vlado.fmf.uni-lj.si/pub/networks/pajek/default.htm

[4] Retrieved from http://vlado.fmf.uni-lj.si/pub/networks/pajek/howto/text2pajek.htm

[5] For instance, in order to prepare the one-mode network of authors connected to authors by comments relationships, we multiplied the transposed two-mode Author-Post network (author name and post ID) with one-mode Post - Comment network (of post ID and comments) (see the formula in the appendix).

[6] Reaction networks RN are transformed so that all the line values are set to 1, which equalizes each reaction in any relationship. Comment networks CN are not transformed this way.

The original networks consisted of many participants, who provided almost no communication within the group (so called "lurkers"). To alleviate computation of the network stability within the positions of the global structure and computing Rand index, we had to reduce the temporal networks (labeled as **CNR1**, **CNR2**, **CNR3**, and **CNR4**; **RNR1, RNR2, RNR3**, and **RNR4**). In both networks in each period, we removed the actors, whose connections with others were not strong enough. The reduced networks were normalized by the logarithmic approach (as the tie values were very skewed) with recoding the link values. It resulted in up to 80 actors in each network, representing group members who were active in communication in the online community, which were easy to navigate while making visualizations of community participants' migrations within the groups. The proper numbers of vertices in each network are provided in Table 2 in Appendix.

**Results**

We start this section with the general statistics of the obtained data. Observing the activity of the group's members, we found an uneven distribution with distinct peaks and falls in activity. We justify the choice of the four periods to which the data were split. After splitting the data in four parts, we look at the content of the most commented posts. Then we look at the global structure of the observed group and check how stable the relations are between the obtained subgroups and within them.

**Members' activity**

In our research frame, the activity in the online community can be seen as the total number of posts, comments, and comments to comments, as well as reactions to posts and comments. Overall, in 7 years, there were 2,591 posts published, which were commented on by 20,709 comments, which, in turn, were connected by an extra 11,005 comments, from 2015. The overall number of reactions of all types is 13,240. The distributions of these data are shown in Figure 1 (the data for 2018 is up to and including January). Comparisons of the distributions of commenting and reacting show that they follow the same trend. The number of comments is usually lower than the number of reactions; however, from the beginning and up to January 2015, there were more reactions than comments. The rise in comments was in 2014 and 2015 (Table 3). In 2016, the number of comments to comments is the highest.

Over 8 years, commenting and reacting activity fluctuated almost every month (Figure 1). We can observe two periods with increased activity: one peak between January – November 2015 (11 months), and another between December 2015 -- May 2016 (6 months). Based on the peaks of activity in the online community, we decided to split our data into 4 periods: two of which are already highlighted, the third from September 2011 -- December 2014 (39 months), and the fourth period from June 2016 -- January 2018 (20 months).

Table 4 (in Appendix) presents some activity statistics for each period. The largest numbers of posts (12,600 and 9,828) are in the first and fourth periods, which are explained by the longer amount of time in these periods. Since the number of posts was not stable during all periods, we decided to normalize the data (dividing the number of posts by the number of months); the most intense periods are the third and the second, with, respectively, 961 and 556 posts written per month, in comparison with 451 on average. The number of group members involved in commenting and providing reactions is also changing during the four periods. As for the actors who commented on the posts and comments, their number (and, in such, diversity) was gradually decreasing from the first to the third period (from 416 to 292 group members), however, in the fourth period, the number increased to 463 members. As for the members providing likes, their number through the four periods increased almost twofold, from 689 in the first period to 1,076 in the fourth one. However, the normalization of these values shows an increased number of community's members commenting to others in the third period (49 actors per month in comparison with 11 on average), as well as those providing reactions to other community's members in the third and second periods (128 and 67 actors per month, respectively, in comparison with 20 on average). The discussions in the group cover many different aspects of sociological profession. Let us go deeper into the description of the second and third periods' content to see what topics of discussions have influenced most the activity in the group social structure.

The second period includes two main peaks in commenting and providing reactions – in March and June 2015 (Figure 1), when there were 41 and 32 posts published, and 1,076 and 1,448 comments received, respectively.

One of the most popular discussions in March 2015, which received 201 comments, is about the results of the study conducted by one of the leading Russian pollster companies devoted to the international relations between Russia and Western counties. The text of the post is as follows (translated by authors; original texts in Russian are provided in Table 5 in Appendix): "*Three polls of the same Levada Center, close in time (November 2014, January 2015, March 2015), the wording differs, at first glance, not very much, and the results show a difference of about 20%. What could have made such an impact, what are the assumptions?*" Some group members believed that the reason for the shift in the results of several waves is in the different wording of questions, while others thought this is because of the changed text on the website. Another popular topic in March 2015, which received 140 comments, is the professional standard of sociologists. The post author problematizes the idea of professional standardization of sociologists in the Ministry of Labor: "*One of the few initiatives when you rejoice in the clumsiness of officials and no doubt vote against it. In a terrible dream, you will not dream of such standardization. I draw attention to the document only to ask: What do you think are the motives for such an initiative and what may be the consequences? Haven't you played enough with the state?*" (Table 5).

In June 2015, the most commented post, which achieved 345 comments, was on the video lecture devoted to the object and subject in sociology written by a creator of this Facebook community, famous Russian professor, a person representing the academician and practical streams of the sociological community. He gave some criticism to the content provided in the video lecture: "*To put the subject and the object on the same line of epistemological work, giving the object a transitive, subordinate status of a transitional entity, is at least strange*" (Table 5). Another post, receiving almost the same (344) number of comments, was published by the same author, and was devoted to the rules for conducting standardized interviews: "*Some people believe that the interviewer should read out the question exactly, not explain or comment on what is written, and avoid any deviations from the questionnaire in every possible way. Others believe that the conversational format, commenting and explaining the meaning of the question, significantly improves the quality of* *answers, so you should stick to the conversational style in a standardized interview. Which point of view is closer to you?*" (Table 5).

In the third period, the peaks of activity in commenting and giving reactions in the community are in January and March 2016 (Figure 1), when 43 and 56 posts and 1,970 and 1,803 comments were posted, respectively. In January 2016, the most popular topic for the discussion was the Crimea poll produced by one of the leading pollsters companies in Russia, having 74 comments. The post was written by the company's director, who emotionally clarified some aspects of Crimea poll, answering to some negative comments concerning the survey process (the wording of questions, conducting a survey during New Year holidays, when the answers could be influenced by the people's mood). Here is the part of the post one the problematic question wording: "*Let me remind you that there were only two questions. There were practically no complaints about the first one. There are many complaints about the second one. We also have them. We did not formulate it, alas. Unlike the first question, when our opinion was taken into account. It did not work out with the second one. Is it a tragedy? Any practicing researcher will say that this is a common situation in their interaction with the client. Perhaps this does not apply to academic scientists, for whom "client" is an abstract concept. For everyone else, it is highly specific, and a compromise on the tools with it is a prerequisite for conducting research. Let us not be hypocritical, colleagues, in such surveys, the research task is most often solved, but the practical task (business, political, corporate, etc.), and not the task of searching for abstract scientific truth. That is why we call ourselves "industrial sociology" and do not claim to be a new word in science*" (Table 5). The comments to this post problematized different aspects: criticized methodological issues of the survey, expressed disbelief to this pollster company because of the governmental roots. However, supportive comments also appeared. In March 2016, the most popular post, which achieved 89 comments, was about the published article discovering the relations between academic sociological science and Russian government. Again, the creator of the community wrote the post and the author of the discussed article did not give any comments.

Observed examples of the actively discussed topics show the issues around which the communication in the group is formed; they all are devoted to the methodological issues, research and analytics, professional skills and knowledge – and thus, show the components of practice and the sphere of interest of this group, seen as a CoP.

**Global structure**

The blockmodeling approach allowed us to extract the global structures of the network of comments **CN** and network of reactions **RN** (Figure 2). As was expected, for both networks the extracted structure can be classified as "core – periphery" type. This structure consists of the two positions – the core of the group members tightly connected to each other (by comments or reactions), and the periphery of the members connected to the members in the core, but ignoring each other. The number of the group members in the core for a **CN** and **RN** are, respectively, 10 and 57. The periphery in both networks includes 808 and 1,482 group members. The periphery also includes a semi-periphery group, which consists of 180 and 326 members, respectively.

**Stability between the positions of the global structure**

To study the stability of the interaction patterns between the positions of the global structure, the blockmodeling approach was applied to the temporal networks of the comments **CN1**, **CN2**, **CN3**, **CN4,** and to the temporal networks of the reactions **RN1, RN2, RN3**, and **RN4**. The obtained structures for both networks in four periods are presented in Figures 3 and 4 (Appendix).

In the reaction temporal network **RN1**, in the first period, a clear division of the core and periphery is found where the core consists of about10 % of all participants. In the second period (network **RN2**), the core decreases, in the third period (**RN3**), it remarkably increases, and finally, in the fourth period, its size returns to the one of the first period. Interestingly, in the third period the structure of the network is changing: the number of those who gave a reaction increased dramatically - 128 reacting actors in comparison with 20 on average (Table 4). The rise in activity probably followed some structural changes. Maybe the periphery did not like what was going on in the community, as seeing what the core was posting. In the second period, there was a part of the periphery that started sharing reactions among each other, but in the following periods, such interactions disappeared.

In the comments temporal network **CN1**, in the first period, six main clusters can be distinguished, where the first cluster consists of only one person, communicating with the whole network. Such an actor is called a "bridging" actor (Kronegger, et al., 2011) and this participant is the leader of this online community. The next two small clusters are semi-periphery – they are partly connected with the core and part of the periphery. The largest cluster is the periphery; however, this cluster could be divided into clusters, with some connections with the semi-periphery and the core, and the true peripheral cluster almost without interactions. In the second period, the peripheral cluster starts constructing tiny groups of people who comment to each other. In the third period, the structure is changing. Part of the semi-periphery starts actively commenting on the core (which also has grown), and another part of the semi-periphery has less activity in communication with each other, but also has some interactions with the core. The peripheral cluster became smaller than in previous periods. In the fourth period, the network **CN4** reverts to the structure similar to that of the first and the second periods.

Overall, all obtained blockmodels for the two networks have a "core-periphery" structural type, so the stability between the positions of the global structure is high. However, in some periods the structure varies: mostly, the size of the core and periphery clusters fluctuates and a cluster of one member, named the "bridging" actor, appears.

**Stability within the positions of the global structure**

To evaluate the stability of the trajectories of individual membership within the positions of the global structure and their change through time, the blockmodeling approach was applied to temporal networks of communication and reaction **CNR1**, **CNR2**, **CNR3**, and **CNR4**; **RNR1, RNR2, RNR3**, and **RNR4**, which were reduced and normalized. For these networks, the blockmodeling approach was applied, which divided the networks into two main clusters of cores and peripheries.

The blockmodeling statistics for the networks of reactions is illustrated in Table 6 (Appendix). The core clusters were formed by 11% of overall network members until the third period when it increased to 15%. In the fourth period, the core decreased to 2%, or just 2 people. The blockmodeling statistics for the networks of comments is illustrated in Table 7 (Appendix). Until the third period, the core clusters were 37% and 38%, but then it decreased to 17%. In the fourth period, the core increased to 44%.

Using the partitions of the clusters to which the actors were assigned by the blockmodeling procedure (core or periphery), we made illustrations of the cluster members' trajectories between these clusters. The terms "incomers" and "outgoers" were proposed to study these kinds of trajectories (Lave and Wenger, 1991), where the first term means the member joining the community, and the second term – the member leaving the group, accordingly.

Figure 5 presents the trajectories of the active part of the community members within the core and periphery in the reaction network over four periods – T1, T2, T3, T4. In each period, there are three clusters to which a member can be assigned to: 1 – the core, 2 – the periphery, NA – people who were not active in that period (had not yet joined the active part of the community or had already left). The main participants of the core seem stable, even if the core became larger in the third period. As for the periphery, there are some members who were consistently present in the active part of the community during all four periods; some members left the active part of the network after the second or third periods. In each period, a large share of the incomers falls into the periphery. Regarding the NA cluster, some members moved there from the periphery after each period and never came back. In some cases, the members moved from the periphery to the NA cluster, and then return. There are also some cases when members left the community after being in the core cluster.

The trajectories for the commenting network are shown in Figure 6. In comparison to the previous network, the commenting network fluctuation is more chaotic. Many more group members entered to the active part of the network in a period, left, and came back again. The participants of the core cluster are changing; however, there is a stable cluster of actors. Interestingly, there is no stability in the participants of the periphery cluster: members come in and leave, and then some of them come back again. A large number of the representatives of the NA cluster only took part in the peripheral cluster in the third period and subsequently left. Some of the community members came to the periphery only in the fourth period.

In order to evaluate the stability of the group members' trajectories, we calculated the modified Rand index (Cugmas, Ferligoj, 2018). For the reaction network, the modified Rand index is 0.3, which is rather low, meaning that the structure in the four periods is not stable. Regarding the core stability, there is only one actor who has been in the same core during all four periods (this is the leader of this community, colored pink). Three actors were stable in the core cluster until the fouth period (green). At the same time, the periphery has some rather stable participants: in all four periods (colored brown), three periods (green) and two periods from the beginning (lilac). The core also has some stable parts for all four periods -- it is colored pink. Overall, in the network based on reactions, the periphery is more stable than the core.

In the commenting network, the modified Rand index is even lower (0.08), which means that the structures in the four periods are not stable at all. However, unlike the reaction network, in the network of comments the core is more stable than the periphery. There are 6 actors, who have been in the core cluster during all four periods (pink). Two actors were stable in the core cluster up to and including the third period (green). The periphery is less stable: there are no community members being stable for all four periods in the same cluster. There is more fluctuation between the clusters: members migrating from the periphery to the core and vice versa.

The stability within the clusters of the global structure is low, according to the modified Rand index (Cugmas, Ferligoj, 2018). However, different patterns were investigated in both networks. In the reaction network the periphery is more stable than the core, while in the comment network the periphery is less stable than the core.

## Discussion

In the Results section, we observed the features of the global structure of the professional online community of Russian sociologists and evaluated the stability between and within its positions. In this section, we would like to discuss how the obtained results can fit to the theoretical model and previous studies.

The global structure of both observed networks can be defined as the "core-periphery" type, in terms of Faust and Wasserman (1992). Though, the comment network is more complex than the reaction network, as besides two main clusters (core and periphery), there is the cluster of semi-periphery. The "core-periphery" structure is a general structural type, but it also can include semi-periphery cluster, which has characteristics of both clusters – core and periphery. The semi-periphery group aspires to get to the core group, which communicates among each other and with the periphery, but it does not have much support from the peripheral group and their communication inside the cluster is not that active as in the core group. The obtained type of structure is in accordance with the other studies of the structure of professional communities (Kronegger et al., 2011; Rykov, 2016; Schenkel et al., 2001). It was shown that the communication in those communities is based on the interaction of the most active participants, while less active participants tend to support and monitor an active group of participants. This is also true for the observed community of sociologists.

To study the stability of the patterns of interactions between the clusters of the global structure (core and periphery) and their change in time, we divided the data into four periods, based on the observation of the group members' activity. In both networks, for each period, the global structure can be characterized as the "core-periphery". However, the number of members in the core, as well as the number of subclusters in both networks, varies according to the peaks of communication and growth of all types of communication (posts, comments, comments to comments, reactions), which the community experienced in the second and third periods. This leads to the appearance of the subclusters among the members of the periphery in both networks. We can assume that the members of the periphery did not fully like the posts and comments written by the core group members, and were interested in communication with each other, forming small subclusters in the second period and larger cluster in the third period; however, this should be a question of a separate inquiry. In the fourth period, the clusters of the global structure return to the characteristics of those in the first period. Another observation relates to the appearance of a cluster of one member - "bridging" actor (Kronegger, et al., 2011), who communicates with the rest of the community.

In both networks, we evaluated the stability of the trajectories of individual membership, within the clusters of the global structure, based on the reduced networks of comments and reaction with the most active members of the community. According to the modified Rand index (Cugmas, Ferligoj, 2018), the structure of three possible trajectories is unstable during all periods in both networks. In comparison to the reaction network, the comments network has more instability in the structure. While comparing the flows of core and periphery members between the clusters, we see the opposite trend - the reaction network has a stable part in the periphery, but there is no such stability in the core, while the commenting network has a stable part in the core, but there is no such stable subcluster in the periphery.

To observe the migration from one cluster to another, we used the types proposed by Lave and Wenger (1991), defining the position and behaviour of participants in the community. Based on the results, we suggest that all those trajectory types are present in our online community, in both networks, based on reactions and comments:

- entries trajectory – for the incomers (from NA), those who just joined the community,
- peripheral trajectory – for the periphery, those who communicate (provide comments and reactions) only to the members of the core,
- internal trajectory – for the core, those who are regular members of the community and actively communicate with all community members,
- borderline trajectory – for the "bridging" actor, who created the community,
- alienation trajectory - for the outgoers (to NA), those who are leaving the community.

The dynamic data shows that the types of trajectories could change their perspectives during time periods. All trajectories have three main perspectives: (1) **foothold**, when the trajectory is stable, (2) **switch,** when participants change cluster (from core to periphery and vice versa), (3) **alienation**, trajectory of leaving the active part of the community, which is visible by means of analysis (as we do not have information if the member really left the Facebook group).

All types of trajectories and perspectives are present in both reaction and commenting networks, and the details are given in Tables 8 and 9. Overall, the only stable trajectory in both networks during all periods is the **borderline trajectory**, which is defined as the trajectory for the community leaders who interact with participants, express themselves in conceptual ideas, and correct any problems of interaction within the community, with the foothold perspective. This trajectory is taken by the "bridging" actor – the leader and creator of the community. As for all other trajectories, they are mixed from one period to another in both networks.

The **entries trajectory** is defined as a trajectory for incomers who just joined the active part of the community. In both networks, most newcomers enter the community as a peripheral member, however for the comment network, the minority of newcomers enter the core from the very beginning. For both networks, the trajectories have all three possible perspectives: foothold, switch, and alienation.

The **peripheral trajectory** is taken by the members of the periphery – those who communicate only with the core by giving comments and reactions. The peripheral trajectory in the network, based on reactions, has two perspectives: foothold and alienation, which means that members of the periphery prefer to stay in the same position or stop giving reactions and leave an active part of the community. In comparison, the members who take the peripheral trajectory in the network, based on comments, can not only stop commenting and leave the active part of the community, but also switch their cluster and enter to the core. In terms of behavior, it means that switching to the core from the periphery is possible only by commenting to the core members.

The **internal trajectory** is for the core members who are actively involved in the community's activities of commenting and giving reactions with each other. In the reaction network, the internal trajectory can be developed in all three perspectives: foothold, switch, and alienation. The comment network mainly has two perspectives: foothold, when the members of the core can stay in the same core , or switch the cluster and relocate to the periphery; however, in some cases the alienation perspective is also possible

The **trajectory of alienation** is the behavior or act of leaving the active part of the community. This trajectory has the same perspectives - switch and alienation - in both reaction and comment networks. People who leave the active part of the community, or become outgoers, could have previously changed their position (from core to periphery, and vice versa) and then stop communication or just leave the active part of the community from their stable (core or periphery) position.

Thus, in this study, we were able to empirically test the model of trajectories proposed by Lave and Wenger (1991), and confirm that all these trajectories can be found in both types of communication, in the online format, by means of reaction and commenting. The proposed trajectories have three main perspectives of development during time periods: staying at the same position (foothold), switching the position (from core to periphery, and vice versa), and alienation (leaving the active part of the community). We found that all these perspectives can characterize the entries trajectories in both networks, which is rather logical: people entering the community may find it interesting and either stay in the periphery, or even enter the core, or leave the cluster from lack of interest. This is also true for those taking an internal position (core of the group) in the reaction network. For the members in the same position, in the comment network, the perspective of alienation is unpopular (however there are some cases). This is a rather interesting observation, meaning that the core members of the online community have chances to either stay at their position or change it to the periphery, but not directly leave the community. Similar perspectives of switching and alienation can be found in both networks for those members having the alienation trajectory. Again, this is also rather logical. The perspective of alienation is also one way of trajectory development for those at the peripheral position. We found that the members of the periphery could also stay at the same position (in the reaction network) and switch to the core position (in the comment network). We can propose that this is due to the difference of relation, but this argument should be checked in further studies. Finally, the borderline trajectory found in both networks is taken by only one community member, who is the creator and leader of this online community.

Based on the previous studies (Sokolov, 2010; Zadorin, Maltseva, 2013), we assumed a possible separation of sociologists in the online community into several groups (such as academics and practitioners, nationally and internationally oriented, etc.). Based on the qualitative observation of the most discussed posts, we noticed that the core reflects the common interests of academics and practitioners. The structural analysis does not support the assumption of division by characteristics, into several groups. Thus, we can conclude that the community under study is able to unite various sociologists from different offline groups and provide a means of communication for those who would like to communicate. Our initial assumption, that this community can be the platform for bringing different people together and forming a joint community, can be confirmed.

Another important aspect is the feature of the "bridging" actor, who appears to be the creator of the community. The role of the community leader is highly important for the community, as he actively participates in the community activity in all periods under study. With some fluctuation, this is the "bridging" actor, who brings stability to the network structure. We can assume that the role of such leaders is extremely important for other professional online communities and CoP.

**Conclusion**

The concept of the community has acquired new features with respect to the development of technologies and the emergence of the Internet. The professional communities adapted to the digital environment, by creating professional groups in social media. By using the optic of CoP, professional communities could also be studied in the online form (Lave Wenger, 1991).

In this study, we analyzed the structural characteristics of the professional online community of Russian sociologists, which exists as a Facebook group, over 7 years (2011 -- 2018) using the perspective of SNA. We were motivated by the fact that the formation of Russian sociology is dramatic and ambiguous for description and reflection, and some previous studies showed the separation between them, in the offline sociological community. We assumed that the online community, proposing flatter and more recursive hierarchies and more horizontal relations, could be the platform for bringing different people together and forming a joint community.

The article shows an original approach to studying communities using a network perspective – defining the global structural type of the community and providing its deep understanding by evaluating the stability of the patterns of interactions between the positions of this structure, as well as the stability of the trajectories of individual membership within these positions.  This approach allows us to discover the insights of the community structure, by identifying not only a complete structure, but also its formation during periods, and detailed analysis of trajectories, in terms of the perspectives of foothold, switch, and alienation. Analysis of the structural changes among periods is valuable for observing the communication among community members. We looked at the migration of the community members from one trajectory to another.

Such analysis could be more disaggregated in periods, for example, using the temporal quantities approach, recently proposed by Batagelj (Batagelj, Maltseva, 2020; Batagelj, 2020); it will remain in the plans for further research. We believe that our methodological approach can be useful for studying different kinds of online communities, including the professional communities and CoP.

**Acknowledgements**

**References**

Barkhatova L. A. (2020) Структурные особенности коммуникации российских социологов: кейс онлайн-сообщества [Structural Features of Russian Sociologists' Communication: an Online Community Case Study]. *Monitoring of Public Opinion: Economic and Social Changes,* 5: 204—221. https://doi.org/10.14515/monitoring.2020.5.1656.

Batagelj V., Mrvar A., Ferligoj A., & Doreian P. (2004). Generalized Blockmodeling with Pajek. *Metodoloski Zvezki*. 1(2): 455-467.

Batagelj V. (2020). On Fractional Approach to Analysis of Linked Networks. *Scientometrics*, 123: 621-633. https://doi.org/10.1007/s11192-020-03383-y

Batagelj V., & Maltseva D. (2020). Temporal Bibliographic Networks. *Journal of Informetrics*, 14(1): 1-14. https://doi.org/10.1016/j.joi.2020.101006

Batigin G., & Deviatko I. (1994). *Тоталитаризм и посттоталитаризм: Социология и власть: эпизоды советской истории* [Totalitarianism and post-totalitarianism: Sociology and power: episodes of Soviet history] (Articles and preparatory materials). IS RAS.

Batigin G., & Gradoselskaya G. (2001). Сетевые взаимосвязи в профессиональном сообществе социологов: методика контент-аналитического исследования биографий [Network connections in the professional community of sociologists: methods of content-analytical study of biographies]. *Sociological journal*, 1: 88-109.

Borgatti S.P., & Everett M.G. (1999). Models of core/ periphery structure. *Social networks,* 21(4): 375-395. https://doi.org/10.1016/S0378-8733(99)00019-2

Bronner M. (2007). *The Meaning of Folklore*. Utah State University Press. Logan.

Chua V., & Madej J., & Wellman B. (2014). *Personal Communities: The World According to Me*. In: The SAGE Handbook of Social Network Analysis. London: SAGE. http://dx.doi.org/10.4135/9781446294413.n8

Cugmas M., & Ferligoj A. (2018). Comparing Two Partitions of Non-Equal Sets of Units. *Metodološki zvezki - Advances in Methodology and Statistics*, 15(1): 1-21. doi

Doktorov B. (2014). *Современная Российская социология: Историко-биографические поиски* [Modern Russian sociology: Historical and biographical searches]. CSEM.

Duncan-Howell J. (2010) Teachers making connections: Online communities as a source of professional learning. *British Journal of Educational Technology*, 41(2): 324–340. https://doi.org/10.1111/j.1467-8535.2009.00953.x

Firsov B. (2012). *История Советской социологии: 1950-1980* [History of Soviet sociology: 1950-1980]. Publishing house of the European University in Saint-Petersburg.

Gorshkov M. (2017). Социология в России: становление и развитие [Sociology in Russia: formation and development]. *Sociological science and social practice*, 2(18): 7-29. https://doi.org/10.19181/snsp.2017.5.2.5147

Goode, W. J. (1957). Community within a community: The professions. *American Sociological Review*, 22, 194–200. https://doi.org/10.2307/2088857

Greenwood E. (1957). Attributes of Profession. *Social Work,* 2 (3): 45-55. https://doi.org/10.1177/0145482X6005400504

Hara N., & Hew K.F. (2007). Knowledge-sharing in an online community of health-care professionals. *Information Technology and People.* 20(3): 235–261. http://dx.doi.org/10.2139/ssrn.1522515

Hur J., & Brush T. (2009). Teacher Participation in Online Communities: Why Do Teachers Want to Participate in Self-Generated Online Communities of K-12 Teachers? *Journal of Research on Technology in Education*, 41(3): 279-303. https://doi.org/10.1080/15391523.2009.10782532

Johnson J.A., Honnold J.A.& Stevens F.P. (2010). Using Social Network Analysis to Enhance Nonprofit Organizational Research Capacity: A Case Study. *Journal of Community Practice*, 18(4): 493-512. https://doi.org/10.1080/10705422.2010.519683

Kronegger L., Ferligoj A., & Doreian P. (2011). On the Dynamics of National Scientific Systems. *Quality & Quantity*, 45(5): 989-1015. https://doi.org/10.1007/s11135-011-9484-3

Korazim-Kőrösy Y., Mizrahi T., Bayne-Smith M. & Garcia M. L. (2014). Professional Determinants in Community Collaborations: Interdisciplinary Comparative Perspectives on Roles and Experiences Among Six Disciplines. *Journal of Community Practice*, 22(1-2): 229-255. https://doi.org/10.1080/10705422.2014.901267

Korpachev A. (2006). Репрезентация сетевых взаимодействий в профессиональном сообществе российских социологов [Representation of network interactions in the professional community of Russian sociologists]. (Bachelor Thesis). Retrieved from Higher School of Economics Dissertations and Theses database https://www.hse.ru/sci/diss/191628148

Kozlova L. (2018). Создание первого академического института социологии – ИКСИ АН СССР, 1968: атмосфера и участвующие субъекты [Creation of the first academic Institute of sociology-ICSI of the USSR Academy of Sciences, 1968: atmosphere and participating subjects]. *Sociological journal*, 3: 117-140. https://doi.org/10.19181/socjour.2018.24.3.5996

Lave, J., & Wenger, E. (1991). *Learning in doing: Social, cognitive, and computational perspectives. Situated learning: Legitimate peripheral participation*. Cambridge University Press.

Louis K., Marks H., & Kruse S. (1996). Teachers' Professional Community in Restructuring Schools. *American Educational Research Journal* 33(4): 757-798. https://doi.org/10.3102/00028312033004757

MacNair R.H. (1996). A Research Methodology for Community Practice. *Journal of Community Practice*, 3(2): 1-19. https://doi.org/10.1300/J125v03n02_01

Maltseva D., Moiseev S., Shirokanova A., Brik T. (2017). Сетевой анализ биографических интервью: возможности и ограничения [Network analysis of biographical interviews: opportunities and limitations]. *Teleskop* 1(121): 29-36.

Maltseva D. & Moiseev S. (2018). Сетевой анализ биографических интервью: кейс Т.Ю. Заславской [Network analysis of biographical interviews: the case of T. I. Zaslavskaya]. *Teleskop* 2(128): 15-24.

Mazina N. (2013). Структурно-семантические особенности репрезентации научных школ и построения сетей (на примере российского социологического сообщества) [Structural and semantic features of the representation of scientific schools and building networks (on the example of the Russian sociological community)]. (Bachelor thesis). Retrieved from Higher School of Economics Dissertations and Theses database https://www.hse.ru/sci/diss/191628148

Maltseva D. (2016). Крымский опрос: анализ дискуссий в онлайн группе «Мануфактура Соцпох" [Crimerian poll: an analysis of discussions in the online group «Manufactura socpoh"]. (Analytical report). Retrieved from Zircon publications http://www.zircon.ru/upload/iblock/e2c/Socpoh_Krymskij_opros.pdf

Nonnecke B., & Preece J. (2003). *Silent Participants: Getting to Know Lurkers Better* // From Usenet to CoWebs. Computer Supported Cooperative Work / Lueg C., Fisher D. London. Springer: 110–132. https://doi.org/ 10.1007/978-1-4471-0057-7_6

Osipov G., Moskvichev L. (2008). *Социология и власть (как это было на самом деле)* [Sociology and power (as it really was)]. Ekonomika.

Podstreshnaya E. (2014). Тема боли в дискурсе о врачебном профессионализме на примере травматологического отделения [How the issue of pain is represented in the discourse on medical professionalism in the case of trauma ward]. *The Journal of Social Policy Studies*, 13(4): 627-642.

Rafaeli S., Ravid G., & Soroka V. (2004). De-lurking in virtual communities: a social communication network approach to measuring the effects of social and cultural capital. *Proceedings of the 37th Annual Hawaii International Conference on System Sciences.*

Renninger K. A., & Shumar W. (2002*).* Building virtual communities: Community building with and for teachers at the Math Forum. In K. A. Renninger, W. Shumar (Eds.), Cambridge University Press (pp. 60-95). https://doi.org/10.1017/CBO9780511606373.008

Reich S., Subrahmanyam K., & Espinoza G. (2012). Friending, IMing, and hanging out face-to-face: overlap in adolescents' online and offline social networks. *Developmental Psychology,* 48(2), 356-368. https://doi.org/10.1037/a0026980

Rheingold, H. (1993). *The virtual community*: Homesteading on the electronic frontier. Reading, Mass: Addison-Wesley Pub. Co.

Rykov Y. (2016). Structure of social networks in virtual communities: comparative analysis of online groups in social media «Vkontakte". (PhD thesis). Retrieved from Higher School of Economics Dissertations and Theses database https://www.hse.ru/sci/diss/191628148

Schenkel, A., Teigland, R., & Borgatti, S. P. (2001). Theorizing Structural Dimensions of Communities of Practice: A Social Network Approach. Paper presented at the annual meeting of the Academy of Management, Washington, D.C.

Schlager M. A., Fusco J., & Schank P. (2002). *Building virtual communities* : Evolution of an online education community of practice. In K. A. Renninger, W. Shumar (Eds.), Cambridge University Press (pp.129-158). https://doi.org/10.1017/CBO9780511606373.010

Simmel, G., & In Wolff, K. H. (1950). *The sociology of Georg Simmel.* Glencoe, Illinois: Free Press.

Sokolov M.M., Bocharov T.Y., Guba K.S. & Safonova M.A. (2010) Проект Институциональная динамика, экономическая адаптация и точки интеллектуального роста в локальном академическом сообществе: Петербургская социология после 1985 года [Institutional dynamics, economic adaptation and points of intellectual growth in the local academic community: St. Petersburg sociology after 1985]. *The journal of sociology and social anthropology*, 8(3): 66-82.

Steinfield Ch., Ellison N., & Lampe C. (2008). Social capital, self-esteem, and use of online social network sites: A longitudinal analysis. *Journal of Applied Development Psychology*, 29(6): 434-445. https://doi.org/10.1016/j.appdev.2008.07.002

Starnes S. (2010). Professional Development for Teachers: Perceptions of Northeast Tennessee Principals. (PhD Thesis). Retrieved from East Tennessee State University Dissertations and Theses database https://dc.etsu.edu/etd/

Subrahmanyam K., Reich S.,Waecher S., & Espinoza G. (2008). Online and offline social networks: Use of social networking sites by emerging adults. *Journal of Applied Developmental Psychology,* 29, 420-433. https://doi.org/10.1016/j.appdev.2008.07.003

Tönnies F. (1957). *Gemeinschaft und Gesellschaft* [Community and society]. East Lansing : Michigan State University Press.

Veatch R. (1972).  Models for Ethical Medicine in a Revolutionary Age. *The Hastings Center Report*, 2(3): 5-7. https://doi.org/10.2307/3560825

Wasserman, S., & Faust, K. (1994). *Structural analysis in the social sciences. Social network analysis: Methods and applications.* Cambridge University Press. https://doi.org/10.1017/CBO9780511815478

Wellman B. (1979). The Community Question: The Intimate Networks of East Yorkers. *The University of Chicago*, 84(5): 1201-1231. https://doi.org/10.1086/226906

Wellman B., Boase J.& Chen W. (2002). The networked nature of community: online and offline. *IT&SOCIETY,* 1(1): 151-165. https://doi.org/10.1111/j.1083-6101.2003.tb00216.x

Wenger E., McDermott R., & Snyder W. (2002). *Cultivating Communities of Practice: A Guide to Managing Knowledge*. HARVARD BUSINESS SCHOOL PRESS. Boston: Massachusetts.

Wirth L. (1938). Urbanism as a way of life. *The American Journal of Sociology* 44(1): 1-24. https://doi.org/10.1086/217913

Zadorin I. & Maltseva D. (2013). Исследование информационной культуры и профессиональной коммуникации в социологическом сообществе [Research of information culture and professional communication in sociological community]. *Teleskop* 3(99): 41-52.

Zadorin I. (2012). Профессиональное сотрудничество социологов. Эссе об одном эксперименте [Professional cooperation among sociologists. Essay about one experiment]. *Sociological journal*, 1: 153-165.

**Appendix**

Table 1. Research on sociology and sociological community in USSR and Russia

| Object of the study | Subject of the study | Reference | Data source | | | | | Method of analysis |
|---|---|---|---|---|---|---|---|---|
| | | | **Documents** | **Interview** | **Survey** | **Social media** | **Bibliographic** | |
| Sociology as science | The history of development of the sociology in USSR and Russia | Firsov, 2012 | + | | | | | Historiographical description |
| | The history of creation of the first academic institute of sociology | Kozlova, 2018 | + | | | | | |
| | History of sociology reborn in the USSR | Moskvichev, Osipov, 2008 | + | | | | | |
| | History of sociology in Russia | Gorshkov, 2017 | + | | | | | |
| Community of sociologists – academy | Individual trajectories of professional career in the academy | Doktorov, 2016 | + | + | | | | Historiographical and biographical description |
| | Professional academic career in collaboration networks | Batigin, Gradoselskaya, 2001; Korpachev, 2006; Mazina, 2013 | | + | | | + | SNA to observe different types of networks |
| | Structural features and meaning of relations underlying the formation of a professional community of sociologists | Maltseva, Moiseev, Shirokanova, Brik, 2017; Maltseva, Moiseev, 2018 | | + | | | | SNA to reveal structural features and meaning of relations underlying the formation community |
| | Structure of the local community of sociologists in St. Petersburg | Sokolov, Bocharov, Guba, Safonova, 2010 | + | | + | | | SNA to reveal the structural patterns of local community in St. Petersburg |
| Community of sociologists – pollsters | Information culture and professional communication | Zadorin, Maltseva, 2013 | | | + | | | Quantitative analysis to observe professional information culture |
| | In-group communication in the professional online community of sociologists about a particular topic | Maltseva, 2016 | | | | + | | SNA to reveal the structure of in-group communication |

Equation 1. Formula for network construction

*Comment network CA:AA(com)*

A two-mode network[7] Actor-Post AP was multiplied with two-mode network Post-Comment PC, it is equally to two-mode network Actor-Comment AC, because Post P was reduced. Then a multiplication of Actor-Comment AC with Comment-Actor CA gave one-mode network[8] of Actors based on comments AA.

$$AP * PC = AC => AC * CA = AA(com)$$

*Reaction network RA:AA(react)*

A two-mode network Actor-Post AP was multiplied with two-mode network Post-Reaction PR, it is equally to two-mode network Actor-Reaction AR, because Post P was reduced. Then a multiplication of Actor-Reaction AR with Reaction-Actor RA gave one-mode network of Actors based on reactions AA.

$$AP * PR = AR => AR * RA = AA(react)$$

Table 2. Number of actors in the reduced networks

| Reduced networks | Periods | Number of actors |
|---|---|---|
| **Comments network CN** | | |
| | CN1 | 79 |
| | CN2 | 72 |
| | CN3 | 94 |
| | CN4 | 75 |
| **Reactions network RN** | | |
| | RN1 | 79 |
| | RN2 | 84 |
| | RN3 | 87 |
| | RN4 | 75 |

---

[7] A two-mode network consists of two sets of units (e. g. people and posts), relation connects the two sets, e. g. posts of people in social media.
[8] A one-mode network consists of one set of nodes(e. g. people).

Table 3. Number of posts, comments to posts, and comments to comments and reactions

| Years | Posts | Comments to posts | Comments to comments | Reactions (all types) |
|---|---|---|---|---|
| **2011** | 189 | 1035 | 0 | 322 |
| **2012** | 479 | 2803 | 0 | 1431 |
| **2013** | 386 | 2860 | 0 | 1884 |
| **2014** | 416 | **4432** | 0 | 2674 |
| **2015** | 367 | **5590** | 387 | **3206** |
| **2016** | 431 | 2406 | **5,954** | 2001 |
| **2017** | 276 | 1367 | 3,936 | 1394 |
| **2018** | 47 | 216 | 728 | 233 |
| **Overall** | 2,591 | 20,709 | 11,005 | 13,145 |

Figure 1. Number of comments and reactions to posts and comments, each month, 2011-2018

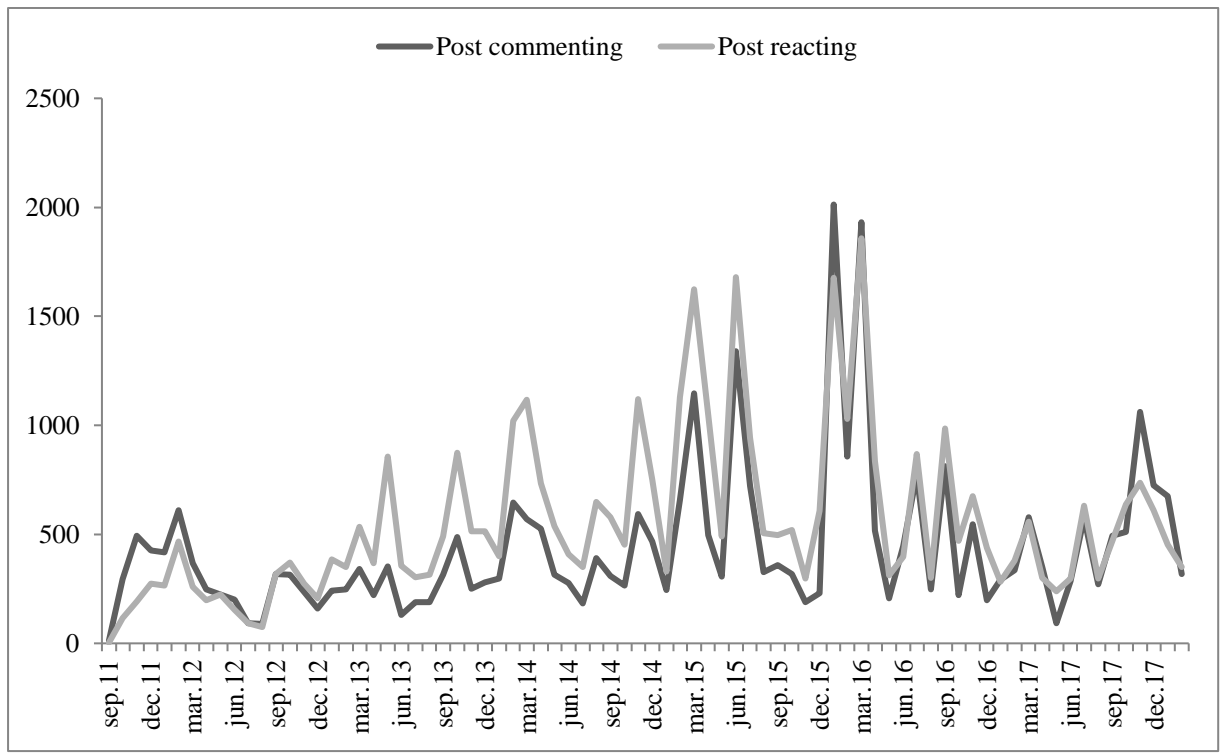

Table 4. Network statistics by 4 periods

| Period | N months | Posts and comments | | actors with comments | | actors with reactions | |
|---|---|---|---|---|---|---|---|
| | | N | Norm | N | Norm | N | Norm |
| **1 (Sept 2011 – Dec 2014)** | 39 | 12,600 | 323 | 416 | 11 | 689 | 18 |
| **2 (Jan – Nov 2015)** | 11 | 6,112 | 556 | 322 | 29 | 740 | **67** |
| **3 (Dec 2015 – May 2016)** | 6 | 5,765 | **961** | 292 | **49** | 769 | **128** |
| **4 (June 2016 – Jan 2018)** | 20 | 9,828 | 491 | 463 | 23 | 1,076 | 54 |
| **Average** | | | 451 | | 11 | | 20 |
| **Total** | 76 | 34,305 | | 818 | | 1,539 | |

Table 5. Texts of mentioned posts and their originals in Russian

| N | Text in English | Original in Russian |
|---|---|---|
| 1 | Two polls of the same Levada Center, close in time (November 2014, January 2015, March 2015), the wording differs, at first glance, not very much, and the results show a difference of about 20%. What could have made such an impact, what are the assumptions? | «Два опроса одного и того же Левада-центра, по времени близко (ноябрь 2014, январь 2015, март 2015), формулировки отличаются, на первый взгляд, не очень сильно, а по результатам разница примерно в 20%. Что могло так повлиять, какие есть предположения? |
| 2 | One of the few initiatives when you rejoice in the clumsiness of officials and no doubt vote against it. In a terrible dream, you will not dream of such standardization. I draw attention to the document only to ask: What do you think are the motives for such an initiative and what may be the consequences? Haven't you played enough with the state? | Одна из немногих инициатив, когда радуешься неповоротливости чиновников и без сомнения проголосуешь против. В страшном сне не приснится такая стандартизация. Обращаю внимание на документ лишь для того, чтобы спросить: Как вы думаете каковы мотивы подобной инициативы и каковы могут быть последствия? Неужели не наигрались с государством? |
| 3 | WHAT IS THE SUBJECT AND OBJECT OF THE STUDY? Victor Vakhshtayn continues his great educational work-he introduces into scientific circulation fundamental concepts that were discarded by the glorious "sociologists" in the backyards of textbooks and dusty manuals. There is nothing more important, from a methodological perspective, than talking about the subject and object of research. Separate bows for this # postnauka and its constant inspirer Ivar Maksutov And as happens in every innovative and from this a little dude action, it was not without incidents and oddities. To put the subject and the object on the same line of | ЧТО ТАКОЕ ПРЕДМЕТ И ОБЪЕКТ ИССЛЕДОВАНИЯ? Victor Vakhshtayn продолжает большое просветительское дело - вводит в научный оборот фундаментальные понятия, отброшенные славными "социологами" на задворки учебников и пыльных методичек. Нет ничего более важного, в методологической перспективе, нежели разговор о предмете и объекте исследования. Отдельные поклоны за это #постнаука и её неизменному вдохновителю Ivar Maksutov И как бывает в каждом новаторском и от этого немного пижонском действии, не обошлось и без казусов и странностей. Ставить предмет и объект на одну линию эпистемологической |

epistemological work, giving the object a transitive, subordinate status of a transitional entity , is at least strange. It turns out that conceptualization is the transformation of an object into an object? Is the turn to things so total and leaves no room for any concept free from the created world? The object as a facet, and not even a polyhedron, but a glass, was left to "the older generation, who inherited epistemological prejudices." Who are you talking about? And why leave room for equivocation and understatement in such an important text? Not a single name, not a single mention of a scientific discussion? In what context is the question posed? Who mentioned such a radical object transitology? Where does the author get the confidence for a linear conceptualization scheme? Questions without answers to which the statement hangs in the air and no mention of freimanalytic games does not save it.

4    Some believe that the interviewer should read out the question exactly, not explain or comment on what is written, and avoid any deviations from the questionnaire in every possible way. Others believe that the conversational format, commenting and explaining the meaning of the question significantly improves the quality of answers, so you should stick to the conversational style in a standardized interview. Which point of view is closer to you?

5    Attention, only on " Socpoh "! A performance ?? "Devil's Advocate", and since some debaters already directly call me "scum", "scum", "executioner", etc. , I will raise the degree myself: Think of me as the devil! If it makes you feel better... And now I will act as my lawyer, and in a separate post, so that those who are too lazy to get into the discussion from the hundredth comment will also have the opportunity to get acquainted with the работы, придавая объекту транзитивный, подчиненный статус переходной сущности - по меньшей мере странно. Оказывается концептуализация - это трансформация объекта в предмет? Неужели поворот к вещам столь тотален и не оставляет места ни для одного свободного от тварного мира понятия? Предмет как грань и даже не многогранника, а стакана, оставлен "старшему поколению, которому достались эпистемологические предрассудки". Это о ком идет речь? И зачем в столь важном тексте оставлять место для экивоков и недосказанности? Ни одного имени, ни одного упоминания научной дискуссии? В каком контексте поставлен вопрос? Кто упоминал о столь радикальной объектной транзитологии? Откуда автор черпает уверенность для линейной схемы концептуализации? Вопросы, без ответов на которые высказывание повисает в воздухе и никакие упоминания о фрейманалитических играх его не спасают.

Среди полстеров присутствуют разные мнения о правилах проведения стандартизированных интервью. Одни считают, что интервьюер должен в точности зачитывать вопрос, не пояснять и не комментировать написанное, всячески избегать каких-либо отклонений от анкеты. Другие полагают, что разговорный формат, комментирование и объяснение смысла вопроса существенно улучшает качество ответов, поэтому следует придерживаться разговорного стиля в стандартизированном интервью. Какая точка зрения вам ближе?

Внимание, только на "Соцпохе"! Выступление ?? "Адвоката дьявола", а т.к. Некоторые дискутанты уже прямо именуют меня "подонком", "мразью", "палачом" и т.п., то повышу градус сам: Считайте меня дьяволом! Если вам так будет легче... А теперь выступлю как свой адвокат, причем в отдельном посте, чтобы с аргументами имели возможность познакомиться и те, кто ленится влезать в дискуссию с сотого комментария:) Итак - заранее извините за arguments:) So-sorry in advance for the volume, it is difficult to write briefly, as the great Chekhov teaches us ... 1. About the wording of the Crimean survey. I recognize their problematic nature in the second case. Let me remind you that there were only two questions. There were practically no complaints about the first one. There are a lot of complaints about the second one. We also have them. We did not formulate it, alas. Unlike the first question, when our opinion was taken into account. It didn't work out with the second one. Is it a tragedy? Any practicing researcher will say that this is a common situation in their interaction with the client. Perhaps this does not apply to academic scientists, for whom "client" is an abstract concept. For everyone else, it is highly specific, and a compromise on the tools with it is a prerequisite for conducting research. Let's not be hypocritical, colleagues, in such surveys, the research task is most often solved, but the task is practical (business, political, corporate, etc.), and not the task of searching for abstract scientific truth. That is why we call ourselves "industrial sociology" and do not claim to be a new word in science. And yet, even in this mode of operation, we often (but not always, and not in this case, of course) manage to simultaneously satisfy our own scientific interest, which goes "in the load" to the interest of the client. In the Crimean survey, such an additional "burden" was excluded due to the lack of time to develop a research program, the extreme lack of time for interviewers, and so on. 2 About sociologists and pollsters, I do not deny in any case that VTsIOM, like other players in the field of "industrial sociology", would be more correct to refer to pollsters. I don't see anything offensive in this attitude. We even considered the idea of creating a proper pollster association, different from a sociological one. But so far, we have decided not to do this - for various reasons, объем, коротко писать трудно, как учит нас великий Чехов... 1. О формулировках крымского опроса. Признаю их проблематичность во втором случае. Напомню, что вопросов было всего два. К первому претензий практически не возникло. Ко второму - претензий много. Есть они и у нас. Не мы его формулировали, увы. В отличие от первого вопроса, когда наше мнение было учтено. Со вторым этого не получилось. Трагедия ли это? Любой практикующий исследователь скажет, что это обычная ситуация в его взаимодействии с клиентом. Возможно, это не касается академических ученых, для которых "клиент" - понятие абстрактное. Для всех остальных - он в высшей степени конкретен, и компромисс по инструментарию с ним - обязательное условие проведения исследования. Давайте не лицемерить, коллеги, в таких опросах всего решается исследовательская задача, но задача практическая (бизнесовая, политическая, корпоративная и т.п.), а не задача поиска абстрактной научной истины. Именно поэтому мы называем себя "индустриальной социологией" и не претендуем на новое слово в науке. И тем не менее даже в таком режиме работы у нас зачастую (но не всегда, и не в этом случае, разумеется) получается попутно удовлетворять собственный научный интерес, который идет "в нагрузку" к интересу клиента. В Крымском опросе такая дополнительная "нагрузка" была исключена ввиду отсутствия времени на разработку программы исследования, крайнего дефицита времени у интервьюеров и проч. 2. О социологах и поллстерах Не отрицаю ни в коем случае, что ВЦИОМ, как и другие игроки поля "индустриальной социологии", правильнее было бы относить к поллстерам. Не вижу в таком отнесении ничего обидного. Мы даже рассматривали идею создания собственно поллстерской ассоциации, отличной от социологической. Но пока решили этого не делать - по разным which can be discussed at length. But not here and not now. And personally, I have never considered myself a sociologist, and I do not consider myself a sociologist, and I have never called myself a sociologist in my 12 years of work at VTsIOM. If anyone is interested, I graduated from the Faculty of Philosophy, and defended my dissertation in political science. I don't see any reason to make a big deal about it. Remember, what faculty did Grushin graduate from? And Levada? And Zaslavskaya? And Poisons?This did not prevent them from becoming the pride of Russian sociology. I do not claim such a status, but I find it stupid and vulgar to bully me on this basis. 3. About the array and the methodological report. The array in the SPSS format has been on our site since January 3. Comments to the notes of @Alexey Kupriyanov will be given at my request by the well-known @Yulia Baskakova, who is much more competent in this topic. We are preparing a methodological report. On the first working week of January, it will definitely appear on our website. We couldn't do it earlier, sorry - everyone is taking a legitimate weekend off, they have every right to do it, because they already had 31.12 and 1.01 in unscheduled work and worries. 4. About the selection 5. About the time of the survey 6. About the tone of the discussion 7. About the future причинам, о которых можно долго говорить. Но не здесь и не сейчас. А уж лично себя я социологом никогда не считал и не считаю, и ни разу за 12 лет работы во ВЦИОМ не называл. Я, если кому интересно, закончил философский факультет, а диссертацию защитил по политическим наукам. Не вижу причин метать по этому поводу громы и молнии. Вспомните-ка, какой факультет закончил Грушин? А Левада? А Заславская? А Ядов?Это не помешало им стать гордостью российской социологии. Я на такой статус не претендую, но и шпынять меня на этом основании нахожу глупым и пошлым. 3. О массиве и методологическом отчете. Массив в формате SPSS висит у нас на сайте с 3 января. Комментарии к заметкам @Алексей Куприянов даст по моей просьбе многим вам известная @Юлия Баскакова, она в этой теме гораздо более компетентна. Методологический отчет готовим. На первой же рабочей неделе января он обязательно появится у нас на сайте. Раньше сделать не смогли, извините - все отгуливают законные выходные, имеют на это полное право, ведь и так у них 31.12 и 1.01 прошли во внеплановых трудах и заботах. 4. О выборке 5. О времени проведения опроса 6. О тоне дискуссии 7. О будущем

Figure 2. Blockmodel of the Comment and Reaction networks **CN** and **RN**

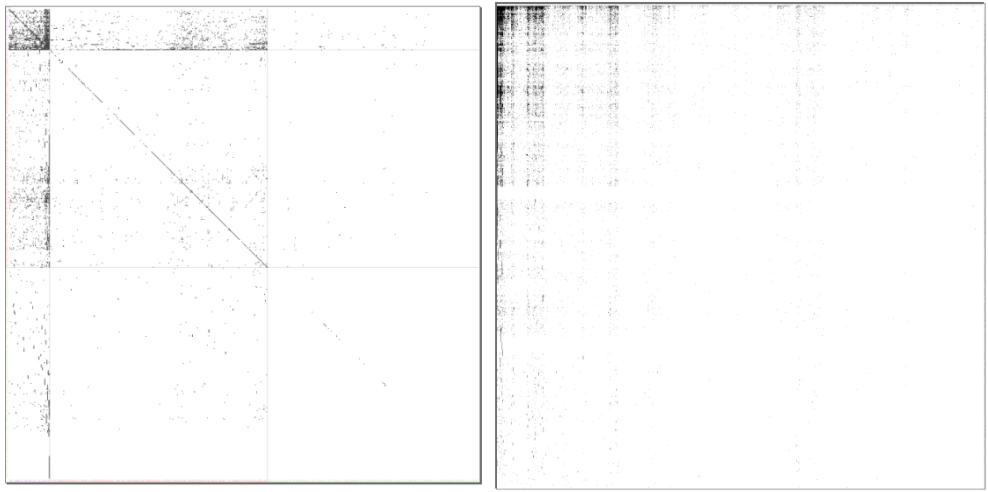

Figure 3. Blockmodels of the commenting networks **CN1**, **CN2**, **CN3**, and **CN4**

CN1                                    CN2

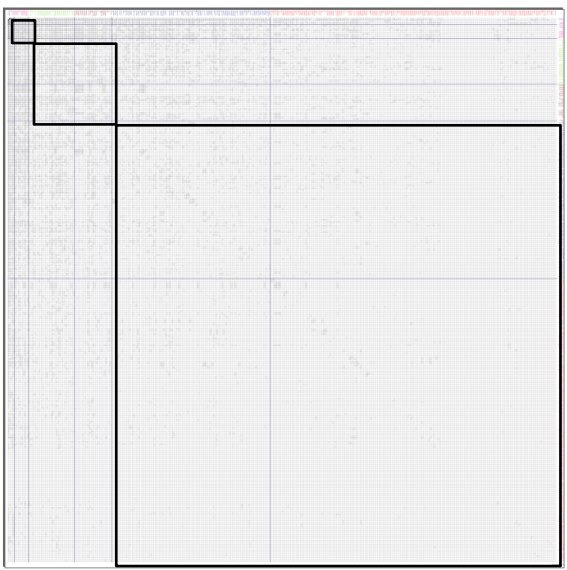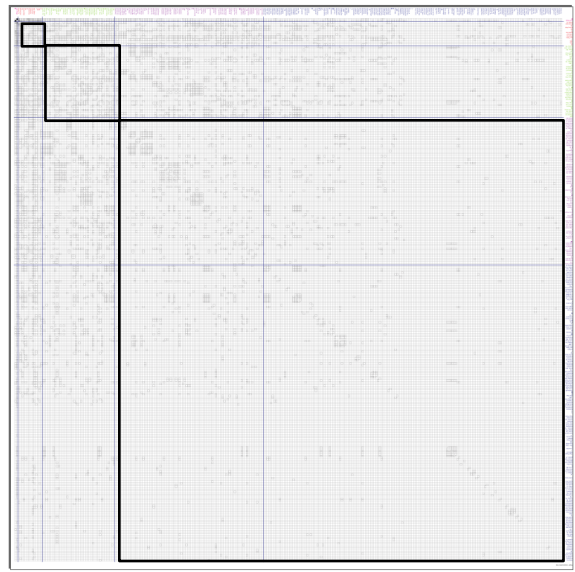

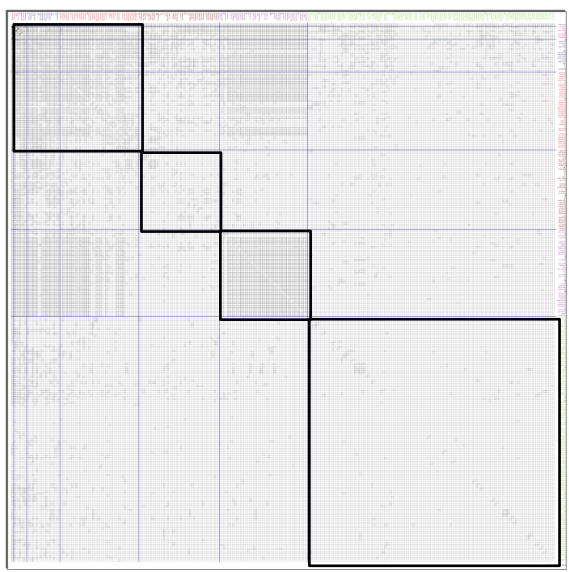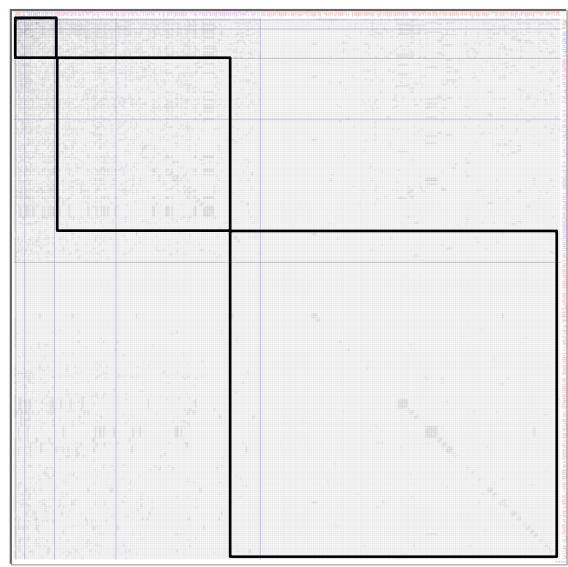

CN3 CN4

Figure 4. Blockmodels of the reaction networks **RN1, RN2, RN3**, and **RN4**

RN1                              RN2

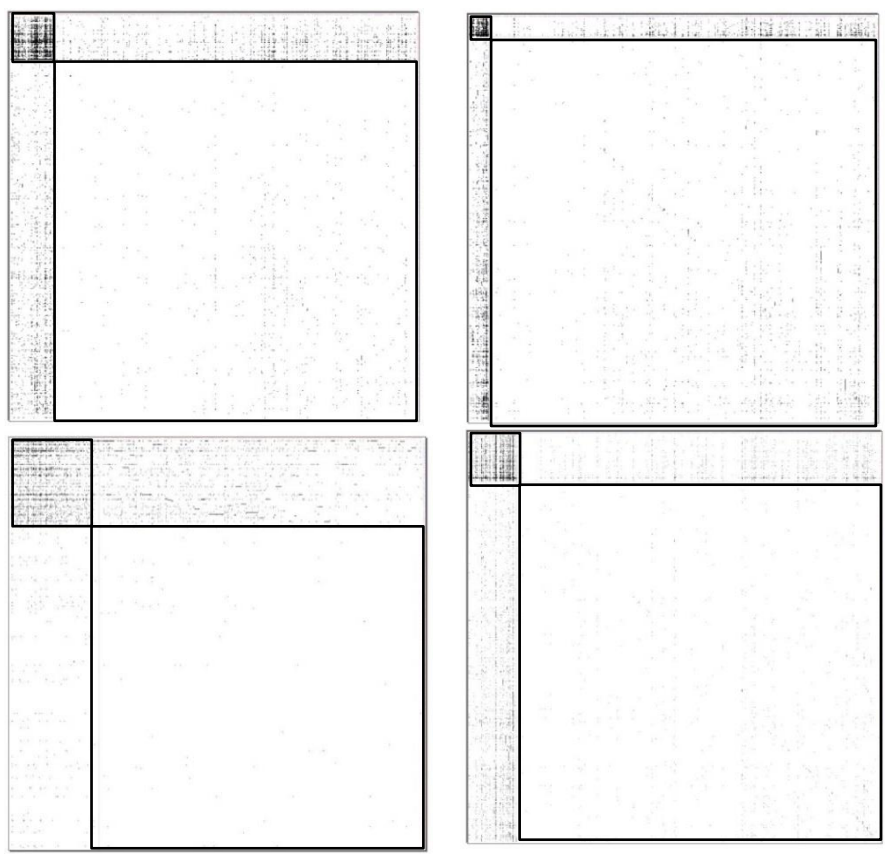

RN3                              RN4

Table 6. Blockmodeling statistics: core and periphery blocks in the reaction network **RN**

|           | *RN1* | | *RN2* | | *RN3* | | *RN4* | |
|-----------|-------|-----|-------|-----|-------|-----|-------|-----|
|           | *N*   | *%* | *N*   | *%* | *N*   | *%* | *N*   | *%* |
| *Core*    | 9     | 11  | 9     | 11  | 13    | 15  | 2     | 2   |
| *Periphery* | 70  | 89  | 75    | 89  | 74    | 85  | 86    | 98  |
| *Overall* | 79    | 100 | 84    | 100 | 87    | 100 | 88    | 100 |

Figure 5. Trajectories within core and periphery in reaction network **RNR**

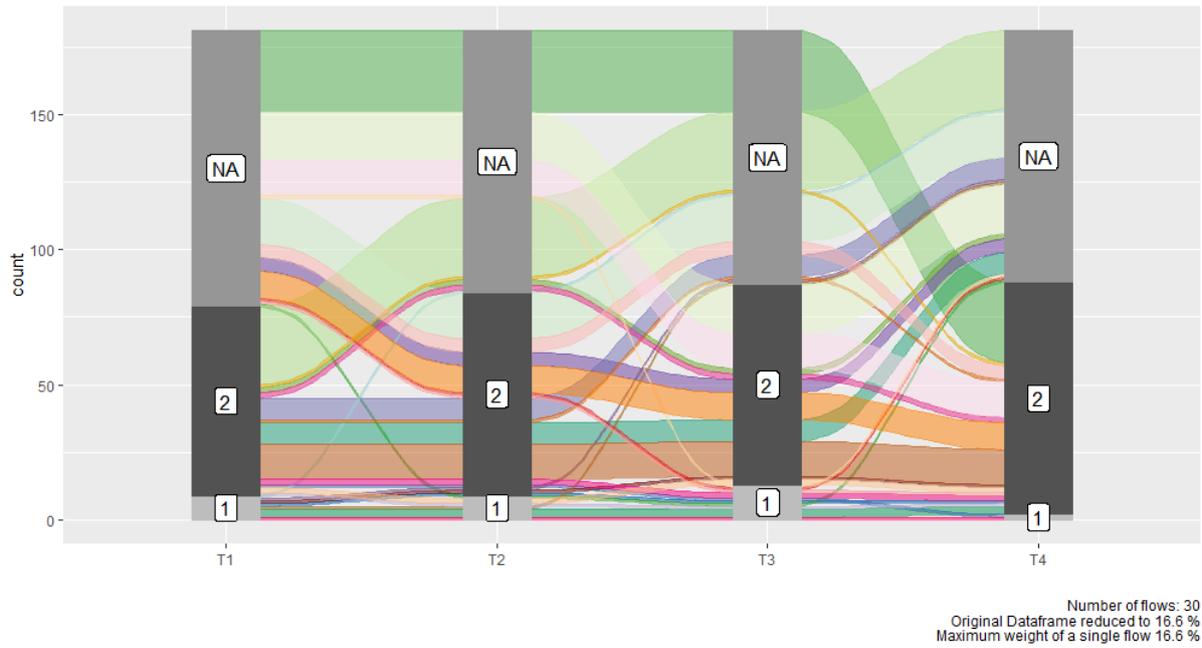

Table 7. Blockmodeling statistics: core and periphery blocks in the commenting network **CN**

|  | **CN1** | | **CN2** | | **CN3** | | **CN4** | |
|---|---|---|---|---|---|---|---|---|
|  | *N* | *%* | *N* | *%* | *N* | *%* | *N* | *%* |
| *Core* | 29 | 37 | 27 | 38 | 16 | 17 | 33 | 44 |
| *Periphery* | 50 | 63 | 45 | 63 | 78 | 83 | 42 | 56 |
| *Overall* | 79 | 100 | 72 | 100 | 94 | 100 | 75 | 100 |

Picture. 7. Trajectories within core and periphery in comment network **CNR**

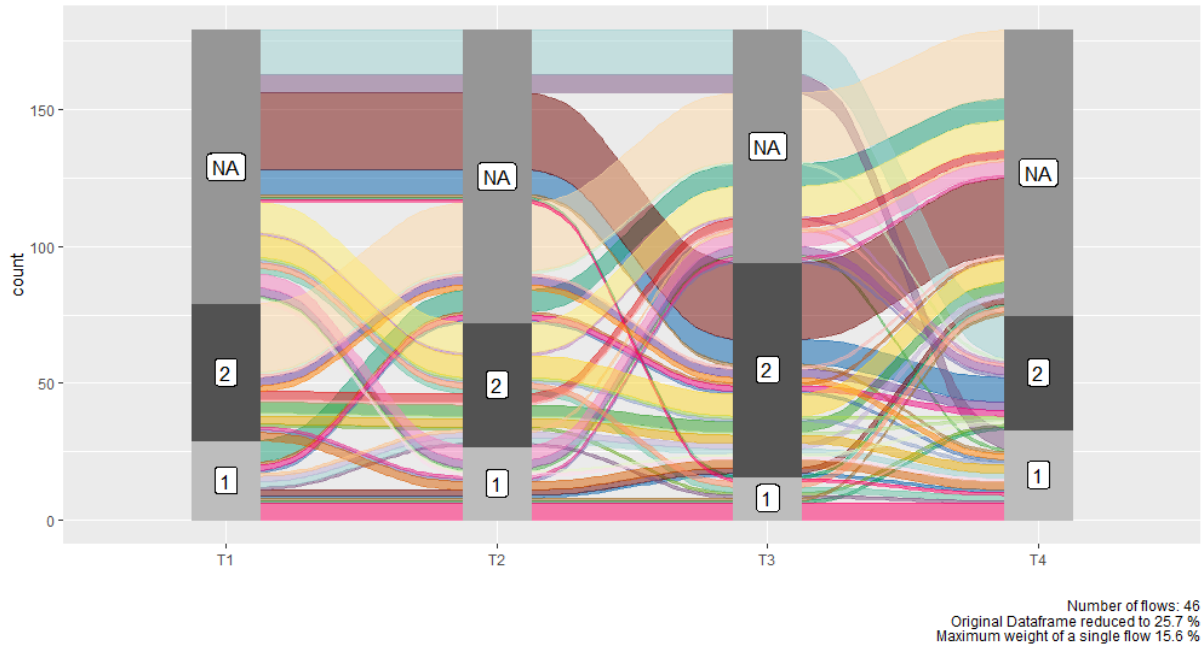

Table 8. Dynamical trajectory types in the reaction network **RNR**

| Trajectories | Foothold | Switch | Alienation |
|---|---|---|---|
| *Peripheral* | + | | + |
| *Entries* | + | + | + |
| *Internal* | + | + | + |
| *Borderline* | + | | |
| *Alienations* | | + | + |

Table 9. Dynamical trajectory types in the comment network **CNR**

| Trajectories | Foothold | Switch | Alienation |
|---|---|---|---|
| *Peripheral* | | + | + |
| *Entries* | + | + | + |
| *Internal* | + | + | |
| *Borderline* | + | | |
| *Alienations* | | + | + |